\documentclass[sigconf]{acmart}

\usepackage{hyperref}
\hypersetup{pdfborder=0 0 0}
\usepackage{cleveref}

\usepackage{graphicx}
\usepackage{tikz}
\usepackage{xcolor}

\usepackage{verbatim} 

\usepackage{caption}
\usepackage{subcaption}

\usepackage{booktabs}
\usepackage{multirow}

\AtBeginDocument{%
  \providecommand\BibTeX{{%
    \normalfont B\kern-0.5em{\scshape i\kern-0.25em b}\kern-0.8em\TeX}}}

\setcopyright{acmlicensed}

\copyrightyear{2024}
\acmYear{2024}
\setcopyright{rightsretained}
\acmConference[WSESE '24]{International Workshop on Methodological Issues with Empirical Studies in Software Engineering}{April 16, 2024}{Lisbon, Portugal}
\acmBooktitle{International Workshop on Methodological Issues with Empirical Studies in Software Engineering (WSESE '24), April 16, 2024, Lisbon, Portugal}
\acmDOI{10.1145/3643664.3648211}
\acmISBN{ 979-8-4007-0567-0/24/04 }

\begin{document}

\title{A Second Look at the Impact of Passive Voice Requirements on Domain Modeling: Bayesian Reanalysis of an Experiment}

\author{Julian Frattini}
\orcid{0000-0003-3995-6125}
\author{Davide Fucci}
\orcid{0000-0002-0679-4361}
\email{{firstname}.{lastname}@bth.se}
\affiliation{
  \institution{Blekinge Institute of Technology}
  \streetaddress{Valhallavägen 1}
  \city{Karlskrona}
  \postcode{37140}
  \country{Sweden}
}

\author{Richard Torkar}
\orcid{0000-0002-0118-8143}
\email{richard.torkar@gu.se}
\affiliation{
  \institution{Chalmers and University of Gothenburg}
  \city{Göteborg}
  \postcode{41756}
  \country{Sweden}}
\affiliation{
  \institution{Stellenbosch Institute for Advanced Study (STIAS)}
  \city{Stellenbosch}
  \country{South Africa}
}

\author{Daniel Mendez}
\orcid{0000-0003-0619-6027}
\email{daniel.mendez@bth.se}
\affiliation{
  \institution{Blekinge Institute of Technology}
  \streetaddress{Valhallavägen 1}
  \city{Karlskrona}
  \postcode{37140}
  \country{Sweden}
}
\affiliation{
  \institution{fortiss GmbH}
  \streetaddress{Guerickestraße 25}
  \city{Munich}
  \postcode{80805}
  \country{Germany}
}

\renewcommand{\shortauthors}{Frattini, et al.}

\begin{abstract}
  The quality of requirements specifications may impact subsequent, dependent software engineering (SE) activities.
  However, empirical evidence of this impact remains scarce and too often superficial as studies abstract from the phenomena under investigation too much.
  Two of these abstractions are caused by the lack of frameworks for causal inference and frequentist methods which reduce complex data to binary results.
  In this study, we aim to demonstrate (1) the use of a causal framework and (2) contrast frequentist methods with more sophisticated Bayesian statistics for causal inference.
  To this end, we reanalyze the only known controlled experiment investigating the impact of passive voice on the subsequent activity of domain modeling.
  We follow a framework for statistical causal inference and employ Bayesian data analysis methods to re-investigate the hypotheses of the original study.
  Our results reveal that the effects observed by the original authors turned out to be much less significant than previously assumed.
  This study supports the recent call to action in SE research to adopt Bayesian data analysis, including causal frameworks and Bayesian statistics, for more sophisticated causal inference.
\end{abstract}




\maketitle

\section{Introduction}
\label{sec:intro}

Requirements specifications serve as input to several subsequent software engineering (SE) activities~\cite{wagner2019status}.
Consequently, the quality of requirements specifications impacts the performance of these dependent activities~\cite{mendez2017naming}.
For example, ambiguous or incomplete requirements specifications may result in incorrect or missing features when implementing the requirements.
Because the cost for remediating these defects scales the longer they remain in the development process~\cite{boehm1988understanding}, organizations are interested in detecting and removing requirements quality defects as soon as possible.

The requirements quality research domain aims to meet this need~\cite{montgomery2022empirical}.
However, while requirements quality research abounds with normative rules about requirements quality~\cite{frattini2022live}, it lacks empirical evidence that supports the relevance of these rules~\cite{frattini2023requirements,montgomery2022empirical}.
Moreover, the few studies contributing empirical evidence are often confounded, too abstract, and their inference reduces complex, context-sensitive data to binary results, for example, through the use of frequentist methods~\cite{lee2011demystify}.
The insufficient quantity and quality of evidence impede the adoption of requirements quality research in practice~\cite{franch2020practitioners}.

With this study, we aim to demonstrate how more sophisticated inference methods than frequentist approaches derive deeper insights from an empirical study and may even revise frequentist claims.
This paper makes the following contributions:

\begin{enumerate}
    \item A recovery of the analysis of one of the only controlled experiments on requirements quality known to us~\cite{femmer2014impact}.
    \item A reanalysis of the hypothesis of this experiment using more sophisticated statistical methods.
\end{enumerate}

\subsection*{Data Availability}
We disclose all supplementary material, including the data, figures, and analysis scripts, in our replication package.\footnote{\url{https://zenodo.org/doi/10.5281/zenodo.10283010}}

\section{Related Work}
\label{sec:related}

\subsection{Requirements Quality}
\label{sec:related:rq}

Requirements quality research is a sub-domain within requirements engineering (RE) research dedicated to the assessment and improvement of requirements artifacts and processes~\cite{montgomery2022empirical}.
Given the importance of RE to the software development life cycle, the quality of its artifacts and processes plays a major role in project success or failure~\cite{mendez2017naming,wagner2019status}.
For requirements artifacts like (systematic) requirements specifications, use cases, user stories, and others~\cite{mendez2015artefact}, a popular concept to identify quality defects is the \textit{requirements quality factor}~\cite{frattini2022live}.
A requirements quality factor is a normative metric that maps a requirements artifact onto some level of quality based on defined criteria~\cite{frattini2022live}.
One commonly researched requirements quality factor is \textit{passive voice}~\cite{femmer2014impact,kof2007treatment}, which associates the use of passive voice in a natural language (NL) requirements sentence with bad quality since it potentially omits the semantic agent of the sentence~\cite{femmer2014impact}.
For example, the requirements specification ``If the settings \textit{are changed}, ...'' obscures the agent of the requirement.
An active formulation of this specification, ``If an administrator \textit{changes} the settings, ...'' makes the agent explicit.

Recent research has identified a major shortcoming of requirements quality factors, namely their relevance~\cite{frattini2023requirements}.
The requirements quality research domain abounds with publications proposing new quality factors and tools to detect violations against them but lacks empirical evidence for the implied causal relationship, i.e., that the violation causes an actual impact on subsequent SE activities~\cite{bano2015addressing}.
A previous literature survey has revealed that among 57 primary studies proposing requirements quality factors, only 40 discuss their impact at all, and of these, only 11 provide some sort of empirical evidence~\cite{frattini2023requirements}.
Without empirical evidence of the impact of a requirements quality factor on subsequent activities, these factors do not reliably identify requirements quality defects that matter.
Practitioners rightfully harbor skepticism toward requirements quality research given this lack of evidence which impedes research adoption in practice~\cite{franch2020practitioners,femmer2018requirements,phalp2007assessing}.

For example, while several sources advise against the use of passive voice as described above~\cite{femmer2017rapid,genova2013framework,kof2007treatment,pohl2016requirements} only two publications known to the authors investigate its actual impact on subsequent activities.
Krisch et al. conducted a document study in which domain experts classified active and passive requirements sentences as either problematic or unproblematic~\cite{krisch2015myth}.
The results indicate that passive voice is generally unproblematic as adjacent text often compensates for the information omitted due to the passive voice.
Femmer et al. conducted a controlled experiment with university students to assess how passive voice in requirements sentences impacts the domain modeling activity~\cite{femmer2014impact}.
The authors conclude that passive voice requirements increase the number of missing associations with statistical significance but not the number of missing actors or domain objects.

\subsection{Inferential Statistics}
\label{sec:related:inferential}

Most statistical methods applied in SE beyond descriptive statistics are limited to frequentist inferential statistics.
These usually take the form of null hypothesis significance testing (NHST), which stratifies the distribution of a dependent response variable by one or more independent variables and compares their mean.
We assume that the popularity of these methods stems from the established guidelines~\cite{wohlin2012experimentation}, the availability of tools to perform them, and their acceptance in the community.

However, frequentist methods like NHST have several shortcomings.
From a research design perspective, they overemphasize the variables involved in an alleged, causal relationship without a systematic approach for addressing confounders~\cite{pearl2016causal}.
From a data analysis perspective, common issues like the multiple-hypothesis problem~\cite{benjamini1995controlling} and the unscientific practice of fishing for significant test results below an arbitrary significance level~\cite{barnes2021statistical} are well-known, yet still occur in practice~\cite{menzies2019bad}.
Moreover, NHST reduces complex, context-sensitive data down to binary answers (i.e., whether there is a significant difference in the distributions' mean or not), which leads to superficial and overly abstracted research results that are void of any uncertainty that the data originally encoded~\cite{furia2019bayesian}.

The recent rise of Bayesian data analysis (BDA) aims to mitigate these shortcomings~\cite{mcelreath2020statistical,lee2011demystify} by (1) embedding inferential statistics in causal reasoning frameworks~\cite{pearl2016causal,siebert2023applications} and (2) applying Bayesian statistics, i.e., encoding the uncertainty of the impact that independent variables have on dependent variables in probability distributions~\cite{mcelreath2020statistical}.
Prior to any data analysis, involved variables and their causal relationship are made explicit.
During the data analysis, explicit prior assumptions are updated in light of the observed data using Bayes' Theorem.
As a result, BDA produces uncertainty-preserving statistical inferences with explicit causal assumptions.
Recently, SE researchers have started to advocate for the adoption of BDA methods~\cite{furia2019bayesian,torkar2020bayesian,furia2022applying} but they still remain to be niche~\cite{siebert2023applications}.

\section{Method}
\label{sec:method}

In this study, we aim to demonstrate how frameworks for causal inference and Bayesian statistics provide more sophisticated insights which reduce issues of drawing inappropriate conclusions from empirical studies.
To this end, we reanalyzed the data of a previous controlled experiment using BDA.
\Cref{sec:method:original:design} presents the design of the original experiment and \Cref{sec:method:original:issues} elaborates on the issues with the experiment.
\Cref{sec:method:reanalysis} then presents the reanalysis performed in the scope of this study.

\subsection{Original Experiment}
\label{sec:method:original}

The original experiment by Femmer et al. aims to understand the impact of passive voice in requirements on domain modeling~\cite{femmer2014impact} by asking the following research questions:

\begin{itemize}
    \item RQ1.1: Is the use of passive sentences in requirements harmful for finding actors?
    \item RQ1.2: Is the use of passive sentences in requirements harmful for identifying domain objects?
    \item RQ1.3: Is the use of passive sentences in requirements harmful for identifying associations?
\end{itemize}

\subsubsection{Design}
\label{sec:method:original:design}

\begin{figure}
    \centering
    \includegraphics[width=0.48\textwidth]{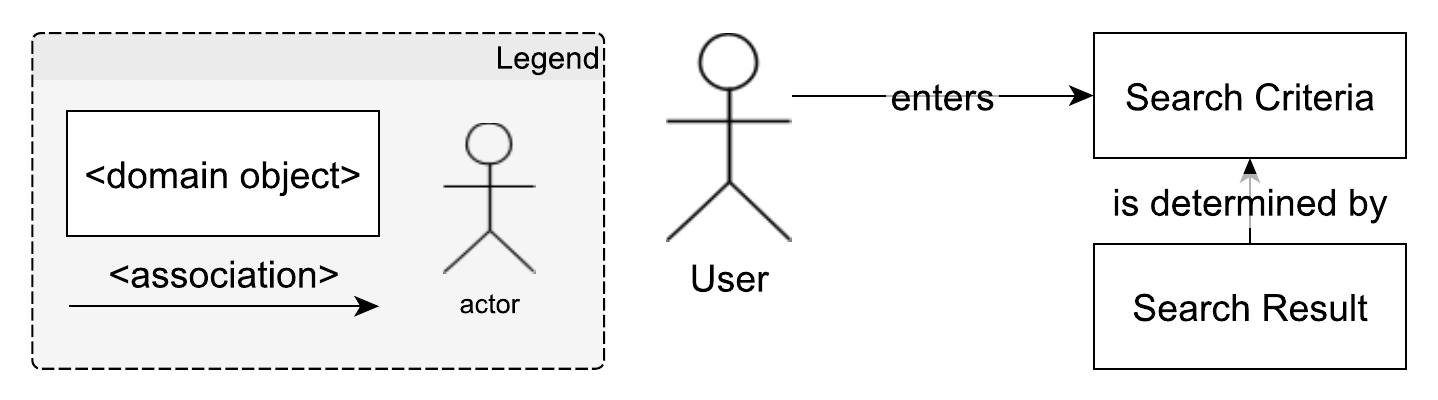}
    \caption{Domain model example}
    \label{fig:domain-model}
\end{figure}

The experimental task was to create a domain model based on a single-sentence NL requirements specification.
The domain model consisted of the following three types of elements:
\textit{actors}, which represent human participants in the requirement, \textit{domain objects}, which represent any non-human entities in the requirement, and \textit{associations}, which connect elements that have a relationship according to the requirement.
\Cref{fig:domain-model} visualizes a domain model for the requirements specification ``The system shall be capable of returning the search results latest 30 seconds after the user has entered the search criteria.''~\cite{femmer2014impact}

The authors of the original study conducted a controlled experiment with independent measures, i.e., every participant is assigned to only one treatment~\cite{wohlin2012experimentation}.
The authors recruited $n_p=15$ participants for the experiment.
The participants consisted of two Bachelor students, eight Master students, four Ph.D. students, and one student with an unknown background.
In addition to the participants' study program, the authors also recorded their age group as well as their industrial and academic experience in SE, RE, and programming on an ordinal scale.

To enable independent measures, seven participants were assigned to the control group (A) and eight to the treatment group (P).
The control group received the requirements specifications in active formulation. 
The treatment group received semantically similar requirements specifications in passive formulation.
For example, the authors transformed the aforementioned active requirements sentence to the following passive formulation for the treatment group: ``The search results \textit{shall be returned} no later 30 seconds after the user has entered the search criteria.''~\cite{femmer2014impact}.

After assessing the general SE and RE knowledge in a quiz, the participants conducted the experimental task.
Every participant received $n_r=7$ requirements specifications, such that the experiment produced $n_p \times n_r = 105$ observations.
The authors then compared the 105 domain models with the sample solution and counted the number of missing actors, domain objects, and associations.
To evaluate the hypotheses implied by the research questions, the authors summed up these numbers for all seven requirements sentences of each participant.
Each participant was associated with a total number of missed actors, domain objects, and associations throughout all seven requirements.
Then, the authors calculated the mean and median number of missing elements for the control and treatment groups and conducted a Mann-Whitney test with a 95\% confidence interval to determine whether there was a statistically significant difference between the two groups.

\begin{table}
    \centering
    \caption{Results of the original study~\cite{femmer2014impact}. P-values indicating a statistically significant difference with $\alpha=0.05$ are prefixed with an asterisk (*)}
    \label{tab:results:original}
    \begin{tabular}{r|cc|cc|ccc}
        \textbf{Element} & \rotatebox[origin=c]{90}{Mean (A)} & \rotatebox[origin=c]{90}{Mean (P)} & \rotatebox[origin=c]{90}{Median (A)} & \rotatebox[origin=c]{90}{Median (P)} & \rotatebox[origin=c]{90}{P-value} & \rotatebox[origin=c]{90}{Conf. Int.} & \rotatebox[origin=c]{90}{Cliff's $\delta$} \\ \hline
        Actors & 0.43 & 1.00 & 0 & 1 & 0.10 & (0; $\infty$) & 0.39 \\
        Objects & 1.29 & 2.00 & 1 & 1 & 0.25 & (-1; $\infty$) & 0.25 \\
        Associations & 4.14 & 7.88 & 3 & 8 & *0.02 & (1; $\infty$) & 0.75 \\ \hline
    \end{tabular}
\end{table}

\Cref{tab:results:original} shows the results of the original study~\cite{femmer2014impact}.
With a significance level of $\alpha=0.05$, the NHST rejects only the null hypothesis implied by RQ1.3 ($p=0.02<\alpha$).
The authors conclude that the use of passive voice does not have a statistically significant impact on the number of actors and domain objects missing from resulting domain models, but it does have an impact on the number of missing associations.

\subsubsection{Issues}
\label{sec:method:original:issues}

The original experiment by Femmer et al.~\cite{femmer2014impact} suffers from at least the following issues.

\paragraph{Issues with reproduction}
The authors originally disclosed their experiment data at \url{http://goo.gl/WlTPE5}, which was forwarded to \url{https://www.in.tum.de/i04/~femmer/data/passives_experiment.zip}.
However, this link does no longer resolve given that institutional websites commonly discontinue hosting resources of members that change their affiliation~\cite{gabelica2022many,winter2022retrospective}.
Thankfully, the authors of the original paper were able to recover the lost replication package~\cite{frattini2023let} and archived it via Zenodo.\footnote{Now available at \url{https://zenodo.org/records/7499290}}
Still, the replication package contains only the study protocol and obtained data, but not the script to reproduce the evaluation.
The lack of reproducibility impedes our goal of comparing methods of statistical inference.

\paragraph{Issues with drawing appropriate conclusions}
The employed research design and analysis risks drawing inappropriate conclusions in two regards.
Firstly, the significance test investigates the isolated impact of passive voice on the three dependent variables.
Possible confounders, like the experience of participants, were recorded but not considered in the evaluation.
Secondly, frequentist NHSTs reduce the data to single, binary results, omitting any uncertainty~\cite{furia2019bayesian} and comparing point estimates, which are unreasonably precise.

\paragraph{Issues selecting an appropriate study design}
The selected experimental design introduced one more potential confounder.
Because the authors of the original study used an \textit{independent measures} design~\cite{wohlin2012experimentation} the evaluation does not account for between-subject variance~\cite{vegas2015crossover}.
In other words: the evaluation does not consider that the observed differences in the dependent variables are caused by the treatment or by other factors like the individual skill of each participant.

\subsection{Reanalysis}
\label{sec:method:reanalysis}

We address the first of the three issues by reproducing the original evaluation and disclosing it for future replication.
For this, we extracted the experimental results from the original study and performed the evaluation according to the information in the manuscript~\cite{femmer2014impact}.
The reproduced evaluation script is contained in our replication package.

To address the second and third issue, we reanalyze the data generated by the experiment using an established framework for causal inference and Bayesian instead of frequentist methods.
The framework allows us to (1) revise and extend the causal assumptions of the original experiment and (2) consider potential confounders in the analysis, while the use of BDA allows us to (3) generate more sophisticated inferences that preserve the uncertainty of the causal influences.

We employ the framework for statistical causal inference that was developed by Siebert~\cite{siebert2023applications}.
This framework is based on Pearl's original model of causal inference~\cite{pearl2016causal} and consists of the three major steps modeling, identification, and estimation.
The following paragraphs briefly summarize each of these steps and are further elaborated in our replication package.
For a gentler introduction to frameworks for statistical causal inference, we refer the interested reader to appropriate literature~\cite{pearl2016causal,siebert2023applications}.
For a gentler introduction to BDA, we refer the interested reader to appropriate textbooks~\cite{mcelreath2020statistical} or descriptive demonstrations of the application of BDA in SE research~\cite{ernst2018bayesian,furia2019bayesian,furia2022applying,torkar2020bayesian}.

\subsubsection{Modeling}

In the first step, we make our causal assumptions of the phenomenon under investigation explicit~\cite{siebert2023applications}.
These causal assumptions are specified in a directed acyclic graph (DAG), in which nodes represent variables and directed edges between them represent assumed causal effects of one variable on another~\cite{elwert2013graphical}.
In our reanalysis, the eligible variables are limited to the variables collected during the original experiment~\cite{femmer2014impact}.

\subsubsection{Identification}

In the second step, we select all variables that form the so-called adjustment set.
Four causal criteria inform this selection and prevent variable bias like colliders or backdoors~\cite{mcelreath2020statistical}, mitigating that non-causal correlations do not influence the causal relation of interest.
The selection of the adjustment set mitigates the second issue mentioned in \Cref{sec:method:original:issues}.

\subsubsection{Estimation}

In the third and final step, we derive a regression model from the adjustment set of eligible variables.
We first select an appropriate probability distribution type to represent each of the three response variables based on the maximum entropy criterion~\cite{jaynes03} and ontological assumptions.
All three variables are whole numbers bounded by the number of expected actors, domain objects, and associations.
Consequently, we model all response variables with Binomial distributions.

We model the parameter $p$---which defines the shape of the Binomial distribution---in dependency of all eligible independent variables, called the predictors.
Each predictor is multiplied with a coefficient that represents the strength and direction of the influence that the predictor has on the response variable.
To begin, we assign an uninformative prior distribution to each of these coefficients, i.e., a normal distribution centered around $\mu=0$ with a standard deviation of $\sigma=1$.
This represents our prior belief of the causal relationship between the predictors and response variables, which are yet unknown.
We confirm the appropriateness of the selected priors via prior predictive checks~\cite{wesner2021choosing}.

The predictors of each response variable consist of the independent variables selected during the identification step.
Further, we include the following variables as predictors: 

\begin{itemize}
    \item Intercept: The global average of missing any element of the domain model. This represents the general challenge of creating a domain model from an NL requirements specification, independent of any predictor values.
    \item Participant-specific intercept: The participant-specific average of missing any element of the domain model. This represents the general skill of a participant.
    \item Requirement-specific intercept: The requirement-specific average of missing any element of the domain model. This represents the general complexity of a requirement.
\end{itemize}

While involving a global intercept is a general best practice~\cite{mcelreath2020statistical}, the two group-specific intercepts retain local variance in the model~\cite{ernst2018bayesian}.
The resulting hierarchical model makes use of partial pooling, which is understood to outperform purely global or local models~\cite{ernst2018bayesian,mcelreath2020statistical}.
The inclusion of a participant-specific intercept mitigates the third issue mentioned in \Cref{sec:method:original:issues}, as it represents between-subject variance in the statistical evaluation.

Given the selected probability distribution and predictors, we train one Bayesian model for each of the three response variables with the experimental data gathered during the original experiment~\cite{femmer2014impact}.
We conduct this step using the \texttt{brms} library~\cite{burkner2017brms} in \textsl{R}.
During the training process, Hamiltonian Monte Carlo Markov Chains update the prior distributions of the predictor coefficients to better reflect the impact of the predictors in light of the observed data~\cite{brooks2011handbook}.
This produces the posterior distributions of the predictor coefficients, which then represent the updated belief of the model about the strength and direction of the influence with which a predictor impacts a response variable.
The standard deviation of each coefficient reflects the uncertainty of the impact of its associated predictor.
This further mitigates the second issue mentioned in \Cref{sec:method:original:issues} by retaining the uncertainty of each impact.

We confirm that the model was trained appropriately by inspecting the Markov Chains~\cite{mcelreath2020statistical} and by performing posterior predictive checks~\cite{wesner2021choosing}.
Finally, we evaluate the trained models by plotting the marginal effects of relevant predictors, mainly the use of passive voice.
The marginal plots show the distribution of the response variable for all levels of the selected predictor while keeping all other predictors at representative levels.
The resulting mean predictions and confidence intervals visualize the difference that the chosen predictor has on the response variable.
This visualization represents the isolated effect of that predictor on the outcome.

\section{Results}
\label{sec:results}

\subsection{Reproduction of the original evaluation}

\Cref{tab:results:reanalysis} shows the strict reproduction of the experimental results using the same frequentist methods as the original study~\cite{femmer2014impact}.
The mean and median values match exactly.
The calculated p-values differ (0.10 vs. 0.19, 0.25 vs. 0.50, 0.02 vs. 0.03), but using the same significance level $\alpha=0.05$ would result in the same hypotheses being rejected (i.e., only the hypothesis implied by RQ1.3).
Similarly, the effect size calculated via Cliff's $\delta$ matches with a margin of 0.07.
Only one extreme end of every confidence interval could not be reproduced.
We assume this to be due to incorrect calculation or reporting in the original study.

\begin{table}
    \centering
    \caption{Results of the strict reproduction}
    \label{tab:results:reanalysis}
    \begin{tabular}{r|cc|cc|ccc}
        \textbf{Element} & \rotatebox[origin=c]{90}{Mean (A)} & \rotatebox[origin=c]{90}{Mean (P)} & \rotatebox[origin=c]{90}{Median (A)} & \rotatebox[origin=c]{90}{Median (P)} & \rotatebox[origin=c]{90}{P-value} & \rotatebox[origin=c]{90}{Conf. Int.} & \rotatebox[origin=c]{90}{Cliff's $\delta$} \\ \hline
        Actors & 0.43 & 1.00 & 0 & 1 & 0.19 & (0; 1) & 0.38 \\
        Objects & 1.29 & 2.00 & 1 & 1 & 0.50 & (-1; 3) & 0.22 \\
        Associations & 4.14 & 7.88 & 3 & 8 & *0.03 & (1; 7) & 0.68 \\ \hline
    \end{tabular}
\end{table}

\subsection{Reanalysis of the data using BDA}

\begin{figure*}
    \centering
    \includegraphics[width=0.9\textwidth]{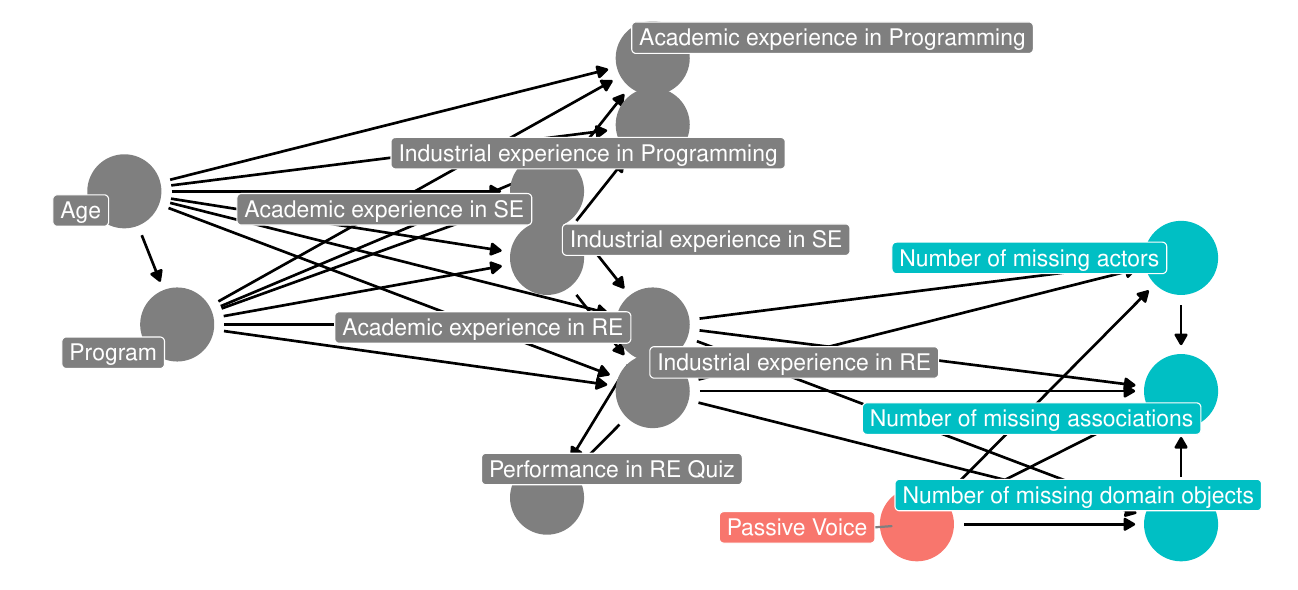}
    \caption{Full DAG visualizing the causal assumptions (red: exposure/main factor, turquoise: response/dependent variables)}
    \label{fig:dag:full}
\end{figure*}

\Cref{fig:dag:full} visualizes the DAG that makes the causal assumptions of the phenomenon under investigation explicit.
The DAG is populated with all variables recorded during the original experiment~\cite{femmer2014impact} and connected with all causal relationships that we assume based on our prior knowledge.
The causal relationships between the main factor (red node) and the three dependent response variables (turquoise nodes) were already assumed in the original study~\cite{femmer2014impact} and are the main relationships of interest.
We assume additional relationships, for example:

\begin{itemize}
    \item Age $\rightarrow$ Program: The older a participant, the more likely it is that they have progressed further in their studies.
    \item Program $\rightarrow$ Academic experience in RE: The more advanced the study program, the higher the academic experience that a student has collected in RE.
    \item Academic/industrial experience in RE $\rightarrow$ number of missing actors/domain objects/associations: The higher the experience in RE, the fewer mistakes a student makes during domain modeling.
    \item Number of missing actors/domain objects $\rightarrow$ Number of missing associations: Missing an actor or domain object leads to missing an association, as one of the two nodes connected through an expected association is unavailable.
\end{itemize}

\begin{figure}
    \centering
    \includegraphics[width=0.48\textwidth]{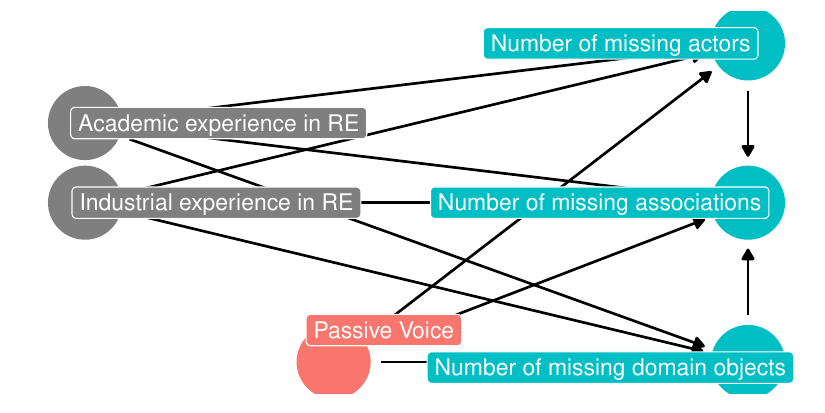}
    \caption{Reduced DAG including all variables eligible for the regression model}
    \label{fig:dag:reduced}
\end{figure}

All other causal assumptions and their justification can be found in our replication package.
\Cref{fig:dag:reduced} visualizes the reduced DAG resulting from the identification step.
This DAG contains only variables included in the adjustment set, i.e., all variables relevant for the causal analysis.
The causal effect of all excluded variables passes through these remaining variables.
Hence, they suffice to model the causal influence on the response variables.

\begin{figure}
    \centering
    \includegraphics[width=0.45\textwidth]{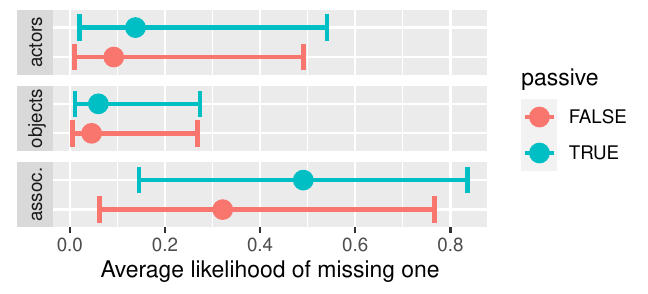}
    \caption{Isolated impact of passive voice on the likelihood of missing an actor, object, or association (``assoc.'')}
    \label{fig:marginal:passive}
\end{figure}

\Cref{fig:marginal:passive} visualizes the marginal effects of the main factor (passive voice) on the three response variables.
All plots show that the use of passive voice slightly raises the mean of the response variable distribution, i.e., the use of passive voice increases the likelihood of missing more actors, domain objects, and associations.
However, the confidence intervals of the main factor overlap in all three cases, meaning that this difference is not significant. 
The chance that the use of passive voice results in equal or even fewer missing actors, domain objects, and even associations remains.

\begin{figure}
    \centering
    \includegraphics[width=0.45\textwidth]{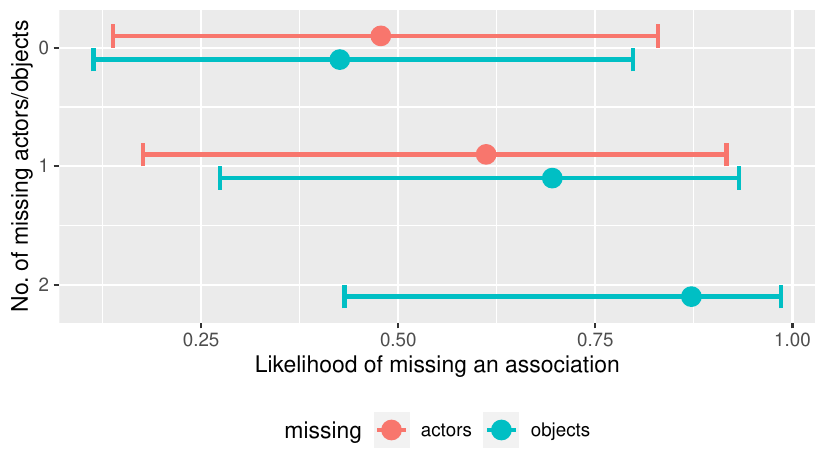}
    \caption{Impact of the number of missing actors and objects on the likelihood of missing an association}
    \label{fig:marginal:transitive}
\end{figure}

\Cref{fig:marginal:transitive} shows the marginal effects of the number of missing actors and missing domain objects on the likelihood of missing an association.
The plot shows that missing an actor or domain model increases the likelihood of missing an association, which confirms the causal assumption represented in our DAG.
The average and confidence interval for the number of missing actors (red in \Cref{fig:marginal:transitive}) is only defined for 0 and 1 because the experiment data did not contain any observation with more than one missing actor per domain model.

\section{Discussion}
\label{sec:discussion}

Finally, we discuss the implications of the results in \Cref{sec:implications} and address remaining threats to validity in \Cref{sec:threats}.

\subsection{Implications}
\label{sec:implications}

Issues of reproduction can be overcome as long as the authors of the original work preserve their replication package.
This encounter supports the observation by Gabelica et al.~\cite{gabelica2022many} and Winter et al.~\cite{winter2022retrospective} that replication packages hosted on institutional websites are prone to become inaccessible over time.
We strongly advise hosting replication packages via services that committed to a long-term retention policy, like Zenodo\footnote{\url{https://zenodo.org/}} or figshare.\footnote{\url{https://figshare.com/}}

More importantly, the reanalysis presented in this study shows that the lack of a framework for causal inference as well as frequentist methods may cause issues with drawing appropriate conclusions.
The results of the reanalysis revealed that the use of passive voice does not have a significant impact on the number of missing associations in resulting domain models as claimed in the original study~\cite{femmer2014impact}.
Instead, the use of a framework for causal inference showed that this impact is confounded by the number of missing actors and domain objects, which also do not experience a significant impact by the main factor of interest.
Additionally, the use of Bayesian statistics highlighted that the remaining difference in the response variables is uncertain and not significantly different.

These insights imply two recommendations for future research.
For research design, the use of an explicit framework for causal inference provides a systematic approach for dealing with potential confounders~\cite{pearl2016causal,furia2022applying}.
For data analysis, the use of Bayesian statistics retains uncertainty and allows transparent inferences from empirical data~\cite{furia2019bayesian,torkar2020bayesian,mcelreath2020statistical}.

\subsection{Threats to validity}
\label{sec:threats}

The reanalysis continues to suffer from threats to validity.
We discuss these according to the classification by Cook et al.~\cite {cook1979quasi}.

\paragraph{Construct validity}
The construct validity suffers from \textit{inadequate preoperational explication of constructs} for all variables concerning experience~\cite{cook1979quasi}.
In the experiment, industrial and academic experience in RE---two of the predictors with an impact on the three response variables---are measured on an ordinal scale with four levels: no experience, up to 6 months, 6 to 12 months, and more than 12 months~\cite{femmer2014impact}.
Whether these variables adequately reflect experience remains questionable.

\paragraph{Internal validity}
The internal validity suffers from potential \textit{confounders}.
The reanalysis could only involve the variables recorded during the original study and was, therefore, constrained to the variables listed in \Cref{fig:dag:full}.
Other variables with a potential causal impact on the response variables---like domain knowledge or prior training in domain modeling---were not available.
The internal validity further suffers from an unknown \textit{interaction with selection} due to the design of the experiment.
Given the independent measures design, each participant was exposed to only one treatment~\cite{wohlin2012experimentation,vegas2015crossover}.
This produced the risk of an interaction effect between the participant and the treatment, i.e., participants of one group could excel with their respective treatment for unknown reasons.

\paragraph{External validity}
The external validity suffers from an \textit{interaction of selection and treatment}, i.e., the experiment participants are potentially not a representative sample of the intended target population.
The study only involved university students of different programs.
Hence, there is no evidence that the conclusions are generalizable to SE practitioners.

\section{Conclusion}
\label{sec:conclusion}

This study reanalyses the only controlled experiment investigating the impact of passive voice in requirements specifications~\cite{femmer2014impact} by employing a framework for statistical causal inference~\cite{siebert2023applications} and using Bayesian in contrast to frequentist data analysis methods~\cite{furia2019bayesian}. 
We could show that the results of the original study are much less significant than suggested by the frequentist analysis and that passive voice has, in consequence, a much smaller impact in the studied context than the original study had assumed.

Needless to say, our aim is not to criticize the original study~\cite{femmer2014impact} itself.
In fact, we would like to acknowledge the authors' contributions to the requirements quality research domain, especially as controlled experiments were, and still are, rare in this domain~\cite{frattini2023requirements}. 
Instead, our intention is to critically reflect upon frequentist analysis that still constitutes the prevalent choice in the empirical software engineering community with little to no attention to its limitations.


Our reanalysis continues to suffer from several threats to validity.
For example, the experimental design made it impossible to identify whether some participants performed particularly well or badly given their assignment to the control or treatment group.
Using a crossover design in which all treatments are applied to all subjects could mitigate this threat~\cite{vegas2015crossover}.

One hope that we associate with our study is to raise awareness of the shortcomings of frequentist analyses, especially when applied as a universal tool. 
We especially hope that our short demonstration, as well as our replication package, will caution fellow SE researchers to use out-of-the-box frequentist approaches and, instead, encourage them to consider Bayesian data analysis approaches~\cite{mcelreath2020statistical}, which include (1) proper frameworks for statistical causal inference~\cite{pearl2016causal,siebert2023applications} and (2) Bayesian statistics~\cite{furia2019bayesian,furia2022applying}.
These approaches ensure that experimental designs are informed by explicit causal assumptions, and their execution produces more sophisticated inferences preserving uncertainty, in turn enriching scientific contributions to be more reflected and insightful.

\begin{acks}
This work was supported by the KKS foundation through the S.E.R.T. Research Profile project at Blekinge Institute of Technology.
We particularly thank Henning Femmer, representing the authors of the original study, for his support and the recovery of the data, which made this reanalysis possible in the first place.
\end{acks}

\bibliographystyle{ACM-Reference-Format}
\bibliography{references}

\end{document}